\documentclass{article}
\usepackage{amsmath,amssymb,subeqnar,graphics,epsfig}

\def\bbbe\mathbb{E}

\def\bbbe{\mathbb{E}}
\def\bbbz{\mathbb{Z}}

\def\ad{\mbox{ad}\,}
\def\diag{\mbox{diag}\,}

\def\tr{\mbox{tr}\,}

\def\im{{\rm Im}\,}

\def\openone{\leavevmode\hbox{\small1\kern-3.3pt\normalsize1}}

\def\otimescomma{\mathop{\otimes}\limits_{'}}

\def\H{{\boldsymbol H}}
\def\q{{\boldsymbol q}}
\def\br{{\boldsymbol r}}

\def\tr{\mbox{tr\,}}
\def\re{\mbox{Re\,}}
\def\im{\mbox{Im\,}}
\def\ad{\mbox{ad\,}}

\allowdisplaybreaks

\begin{document}
\begin{center}

{\Large \bf Selected Aspects of Soliton Theory\\
Constant boundary conditions}

\bigskip
{\bf V. S. Gerdjikov}

\bigskip

{\sl Institute for Nuclear Research and Nuclear Energy\\
1784 Sofia, Bulgaria}

\bigskip

\end{center}

\begin{abstract}
Brief review of the methods for solving the multicomponent nonlinear
Schr\"odinger (MNLS) equations and analysis of their Hamiltonian
structures is given.  Main attention is paid to the MNLS related to
the $C.II $- and $D.III $-types symmetric spaces with nonvanishing
(constant) boundary conditions. The spectral properties of their Lax
operators are described. The derivation of the trace identities is
outlined. The involutivity of their integrals of motion is proved using
the method of the classical $R $-matrix.

\end{abstract}

\section{Introduction} \label{intro}

The integrability of the well known (scalar) NLS eq.:
\begin{equation}\label{eq:3.2}
i q_t + q_{xx} +2 |q(x,t)|^2 q(x,t) =0,
\end{equation}
was discovered by Zakharov and Shabat in their pioneer paper
\cite{ZS*71} which strongly stimulated the search of other
important integrable nonlinear evolution equations (NLEE).

Soon after \cite{ZS*71} Zakharov and Shabat proved the integrability and
physical importance of the NLS with constant boundary conditions
\cite{ZS*73} (CBC):
\begin{equation}\label{eq:3.2c}
i q_t + q_{xx} -2(|q(x,t)|^2 -\rho ^2 ) q(x,t) =0,
\qquad \lim_{x\to\pm\infty } q(x,t)=q_\pm,
\end{equation}
where the asympotic values $q_\pm $ satisfy $|q_\pm|^2=\rho ^2 $.
Note the sign difference in the qubic nonlinearity as well as the
additional term with the chemical potential. These changes are important
from both physical and mathematical point of view.

Both versions of the NLS eq. served as paradigms on which the ISM  was
developed \cite{ZS*71,ZS*73,ZMNP}. They served also as a tool for the
developement of the quantum ISM, see the review paper \cite{4}.

The simplest nontrivial multicomponent generalizations of NLS is
the vector NLS eq. known also as the Manakov model \cite{Man*74a}:
\begin{equation}\label{eq:3.3}
i \vec{q}_t + \vec{q}_{xx} + 2 (\vec{q}{\,}^\dag \vec{q}) \vec{q}(x,t)=0,
\end{equation}
where $\vec{q}(x,t) $ is an $n $-component complex-valued vector tending to
zero fast enough for $x\to\pm\infty  $. The CBC version of the vector
NLS equations (especially the one with $n=2 $)
\begin{equation}\label{eq:3.3cbc}
i \vec{q}_t + \vec{q}_{xx} - 2 \left( ({q}^\dag \vec{q})-\rho ^2\right)
\vec{q}(x,t)=0,
\end{equation}
also finds applications in nonlinear optics, plasma physics etc.
Here $\lim_{x\to\pm\infty }\vec{q}=\vec{q}_\pm $, $\vec{q}_-=U_0\vec{q}_+
$ where $U_0 $  is constant unitary matrix.

The close relations between the sectional curvatures of the symmetric
spaces and the interaction constants of the integrable multicomponent NLS
(MNLS) was discovered in \cite{ForKu*83} for vanishing boindary
conditions (VBC). The basic ideas for constructing their soliton solutions
via the dressing Zakharov-Shabat method were formulated in \cite{Za*Mi}.
These ideas and results were developed in a number of more recent
monographs \cite{APT} and papers, see
\cite{11,TMF99,G,vgn2,GGIK,Varna04,GGK-04,Ivanov} and the numerous
references therein.

However the properties of the MNLS with CBC have substantial differences
as compare to the VBC case. This is due to the facts that: i)~the
corresponding phase space $\mathcal{M} $ spanned by the allowed potentials
is nonlinear; ii)~the relevant Lax operators may have rather involved
spectral properties. Both these facts, noticed long ago
\cite{12,LOMI131,FaTa} (see also \cite{Konotop,Dokt}) explain why MNLS
with CBC have not been so well studied.

In the next Section 2 we provide the  necessary facts from the theory of
MNLS and the relevant symmetric spaces \cite{ForKu*83} for vanishing
boundary conditions. In Section 3 we reformulate these results for CBC. In
particular we describe the spectral properties of MNLS with CBC for
several types of symmetric spaces. In Section 4 we detail two examples of
MNLS related to the algebras $sp(4) $ and $so(8) $. In the last Section 5
we concentrate on the Hamiltonian properties and their classical $R
$-matrix formulation.  In particular we outline how the integrals of
motion from the fundamental series can be regularized so that they become
regular functionals on the non-linear phase space $\mathcal{M} $.

\section{MNLS with vanishing BC}\label{sec:2}

Equations (\ref{eq:3.2}) and (\ref{eq:3.3}) are particular cases
of the matrix NLS eq. which is obtained from the system:
\begin{eqnarray}\label{eq:3.1}
i \q_t + \q_{xx} + 2 \q\br\q(x,t) =0,\nonumber\\
-i \br_t + \br_{xx} + 2 \br\q\br(x,t) =0.
\end{eqnarray}
with appropriate reductions (involution). Here  $\q(x,t) $ and
$\br^T(x,t) $ are $n\times m $ matrix-valued functions of $x $ and
$t $ with $n>1 $, $m>1 $ which are smooth enough and tend to zero
fast enough for $x\to\pm\infty  $. The best known involution
compatible with the evolution of (\ref{eq:3.1}) is
\begin{equation}\label{eq:3.5}
\br=B_- \q^\dag B_+^{-1}, \qquad B_\pm=\diag ( \epsilon _1^\pm,\dots
,\epsilon _m^\pm ), \qquad (\epsilon _s^\pm)^2=1.
\end{equation}
and the corresponding MNLS equation is of the form:
\begin{equation}\label{eq:3.6}
i \q_t + \q_{xx} + 2 \q B_- \q^\dag B_+^{-1} \q(x,t) =0
\end{equation}
For $n=m=1 $ and $r=\epsilon q^* $ the system (\ref{eq:3.1}) goes
into the scalar NLS equation; for $m=1$ and $n>1$ and with
appropriate choice of the involution (\ref{eq:3.5}) eq.
(\ref{eq:3.1}) can be transferred  into the Manakov model or into
eq. (\ref{eq:3.6}).

The MNLS (\ref{eq:3.6}) is known to be closely related to the
symmetric spaces \cite{ForKu*83}. All these versions of NLS are
solvable by applying the ISM to a generalization of the
Zakharov-Shabat system of the form:
\begin{eqnarray}\label{eq:4.1}
L\psi &\equiv & \left( i {d \over d x } + Q(x,t) - \lambda
J\right)
\psi (x,\lambda ) =0, \\
\label{eq:4.1M} M\psi &\equiv & \left(i {d \over d t} +
V_0(x,t) +\lambda V_1(x,t) -
2\lambda^2 J\right) \psi (x,\lambda ) =0, \\
\label{eq:Q}
Q(x,t) &=& \left( \begin{array}{cc} 0 & \q(x) \\ \br(x) & 0
\end{array} \right), \qquad J = \left( \begin{array}{cc} \openone
& 0 \\ 0 & -\openone \end{array} \right),
\end{eqnarray}
where $Q(x,t) $ and $J $ are $(n+m)\times (n+m) $ matrices with
compatible block structure and $V_0(x,t) $, $V_1(x,t) $ are
expressed in terms of $Q $ and its $x $-derivative:
\begin{equation}\label{eq:4.1V}
V_1(x,t) =2Q(x,t), \qquad V_0(x,t) = -[Q, \ad_J^{-1}Q] +
2i \ad_J^{-1}Q_x.
\end{equation}

An effective tool to obtain new versions of MNLS type equations
is the reduction group introduced by  Mikhailov \cite{2}. It allows one to
impose algebraic constraints on $Q(x,t) $ which are automatically
compatible with the evolution. For example, the involution (\ref{eq:3.5})
(or $\bbbz_2$-reduction) can be written as:
\begin{equation}\label{eq:4.2}
B U^\dag (x,t,\lambda ^*) B^{-1} = U(x,t,\lambda ), \qquad
B=\left( \begin{array}{cc} B_+ & 0 \\ 0 & B_- \end{array} \right),
\end{equation}
where $B$ is an  automorphism of $\mathfrak{g} $, i.e. matrix such that
$B^2=\openone $, $[J,B]=0 $  and:
\begin{equation}\label{eq:11a}
U(x,t,\lambda ) = Q(x,t) -\lambda J.
\end{equation}
Reductions leading to new types of MNLS  systems are demonstrated
in \cite{Varna04,manev04}.

The spectral properties of the Lax operators (\ref{eq:4.1}) as well as the
basic properties of the MNLS equations for the $A $-type symmetric
spaces are analyzed in \cite{APT,Varna04,TMF99}. Our aim below is to
extend these results to the MNLS equations with CBC for the $C $- and $D
$-type symmetric spaces.

\section{MNLS with CBC}\label{sec:3}

Though similar in form, the MNLS with CBC require a number of important
changes. These changes are based on two constraints which ensure:
i)~regular behaviour of the solutions for $t\to\pm\infty  $; in other
words we want to avoid strong oscillations for large times; ii)~require
that the spectrum of the two asymptotic operators $L_\pm = id/dx +
U_\pm(\lambda ) $ have the same spectrum. Here
\begin{equation}\label{eq:U_pm}
U(x,t,\lambda )=Q(x,t)-\lambda J, \qquad U_\pm(\lambda ) \equiv
\lim_{x\to\pm\infty } U(x,t,\lambda ) = Q_\pm -\lambda J.
\end{equation}

The first requirement can be satisfied by regularizing the MNLS, i.e.
by conveniently adding linear in $\q $ terms, see eq.  (\ref{eq:1}).
The corresponding regularized  MNLS have the form:
\begin{equation}\label{eq:1}
i\q_t + \q_{xx} - 2 \q\q^\dag \q + \q\mu  + \bar{\mu }\q =0,
\end{equation}
with the boundary conditions
\begin{equation}\label{eq:2}
\lim_{x\to\pm\infty } \q(x,t)=\q_\pm, \qquad \mu =\q_+^\dag \q_+ =
\q_-^\dag \q_- , \qquad \overline{\mu} = \q_+ \q_+^\dag  = \q_- \q_-^\dag
, \end{equation}

The second requirement will be satisfied limiting values $\q_\pm $ must
satisfy also
\begin{equation}\label{eq:4}
Q_+^2 = Q_-^2.
\end{equation}
which ensures that $U_+(\lambda ) $ and $U_-(\lambda ) $ have the same
sets of eigenvalues.

The $M $-operators of the MNLS with CBC (\ref{eq:1}) are given by
(\ref{eq:4.1M}) with
\begin{eqnarray}
V_0(x,t)=- [Q, \ad_J^{-1}Q] + 2i  \mbox{ad}_J^{-1}Q_x(x,t) +[Q_\pm ,
\ad_J^{-1}Q_\pm ].
\end{eqnarray}
The additional terms, as compare to (\ref{eq:4.1V}) ensure the regular
behavior of the solutions for large $t $.

Lax operators of the form (\ref{eq:4.1}) can be associated  with each
of the symmetric spaces described below. They are
defined by specifying the simple Lie algebra $\mathfrak{g} $ (having
typical representation in matrices $N\times N$, $N=s+s' $) and $J $:

\begin{description}

\item [A.II] $\mathfrak{g}\simeq A_{N-1}\equiv sl(N) $, $J=H_{\vec{a}} $,
where the vector $\vec{a} $ in the root space $\bbbe^{r} $ dual to $J $ is
given by $\vec{a}=\sum_{k=1}^{s} e_k -\sum_{k=s+1}^{N} e_k $;

The next two cases require that $s=s'=r $ and that $N=2r $ is even.

\item [C.II] $\mathfrak{g}\simeq C_{r} \equiv sp(2r)$, $J=H_{\vec{a}} $,
where the vector $\vec{a} $ in the root space $\bbbe^{r} $ dual to $J $ is
given by $\vec{a}=\sum_{k=1}^{r} e_k $;

\item [D.III] $\mathfrak{g}\simeq D_{r} \equiv so(2r) $, $J=H_{\vec{a}} $,
where the vector $\vec{a} $ in the root space $\bbbe^{r} $ dual to $J $ is
given by $\vec{a}=\sum_{k=1}^{r} e_k $.
\end{description}

\begin{figure}
\centerline{\includegraphics[width=5.5cm,height=5.5cm,clip]%
{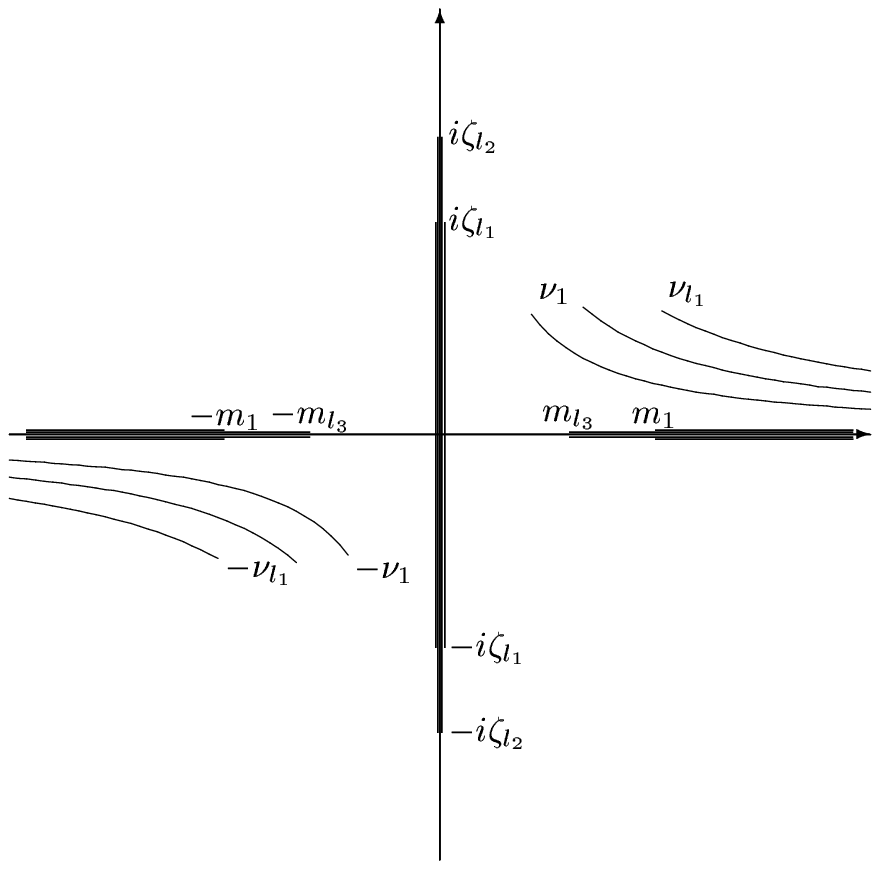}\qquad \includegraphics[width=5.5cm,height=5.5cm,clip]%
{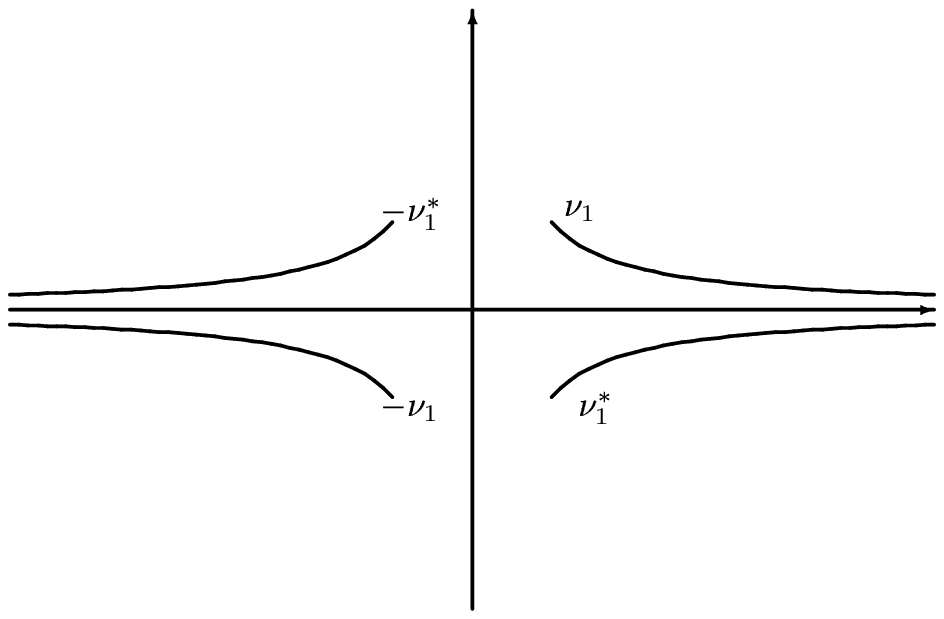}}
\caption{\label{fig:1}
Left panel: the continuous spectrum of $L $, generic case; Right
panel: the continuous spectrum of the $sp(4) $ and  $so(8) $  MNLS with
CBC for $D<0 $; the only difference is that while the multiplicity
of the spectra of $sp(4) $ is 2 the one for $so(8) $ is 4.}
\end{figure}

The spectrum of the asymptotic operators $L_\pm $ is purely continuous and
is determined by the the eigenvalues of $Q_\pm $ which generically may be
arbitrary complex numbers. The spectra of $A $-type symmetric spaces were
described in \cite{LOMI131}. For $C.II $- and $D.III$-type  symmetric
spaces the spectra may consist of the following types of branches, see the
left panel of fig.~\ref{fig:1}.

\begin{description}

\item[a) $\nu _k\neq \pm \nu _k^* $, $k=1,\dots, l_1 $] -- two branches of
two-fold spectrum filling up the hyperbola's arcs $\re \lambda\, \im
\lambda = \re \nu _k \im \nu _k$ on which $|\re \lambda |\geq |\re \nu
_k|$;

\item[b) $\nu _{l_1+k} =- \nu _{l_1+k}^* = i\zeta _k $, $k=1,\dots, l_2 $]
-- two branches of two-fold spectrum filling up the real axis and
the segment $|\im \lambda| \leq |\zeta  _k | $ of the imaginary axis;

\item[c) $\nu _{l_1+l_2+k} =\nu _{l_1+l_2+k}^* = m _k $, $k=1,\dots,
l_3=r-l_1-l_2 +1$] --  two branches of two-fold spectrum filling up the
segments $|\re \lambda| \geq |m _k |$ of the real axis;

\end{description}

In order to proceed with the $C_r $- and $D_r $-types symmetric spaces we
will need to introduce their definitions and Cartan-Weyl basis in the
typical representations. In what follows we will define the Lie algebra
$\mathfrak{g} $ by:
\begin{equation}\label{eq:Lie}
\mathfrak{g} \equiv \left\{ X\colon X+S_0 X^TS_0^{-1} =0 \right\},
\end{equation}
where
\begin{eqnarray}\label{eq:S0sp}
S_0 &=& \sum_{s=1}^{r} (-1)^{s+1} \left( E_{s\bar{s}} - E_{\bar{s}s}
\right), \qquad \mbox{for} \quad \mathfrak{g}\simeq sp(2r), \\
\label{eq:S0so}
S_0 &=& \sum_{s=1}^{r} (-1)^{s+1} \left( E_{s\bar{s}} + E_{\bar{s}s}
\right), \qquad \mbox{for} \quad \mathfrak{g}\simeq so(2r).
\end{eqnarray}
Here $\bar{s} =2r+1-s $ and $E_{ks} $ are $2r\times 2r $ matrices defined
by $(E_{ks})_{jl} =\delta _{kj}\delta _{sl} $. Note that $S_0^2 =\epsilon
_0\openone  $, where $\epsilon _0=-1 $ for $sp(2r) $ and $\epsilon _0=1 $
for $so(2r) $.

By $e_k $, $k=1,\dots ,r $ we denote the  vectors forming an
orthonormal basis in the root spaces $\bbbe^r$  for both types of
algebras. With the above definitions of $\mathfrak{g} $  their Cartan
generators $H_k $ dual to $e_k $ are diagonal and given by:
\begin{equation}\label{eq:Hk}
H_{k} = E_{kk} - E_{\bar{k}\bar{k}},
\end{equation}
As we already mentioned, the element $J $ dual to $\vec{a} $ is
$J=\sum_{k=1}^{r}H_k $. Using $J $ we can split the set of positive roots
into two subsets $\Delta ^+ = \Delta ^+_0 \cup \Delta ^+_1  $ which for
$\mathfrak{g}\simeq sp(2r) $ are:
\begin{eqnarray}\label{eq:d1-sp}
\Delta ^+_0 &=& \{ e_i - e_j \}, \qquad
\Delta ^+_1 = \{2e_i, \quad  e_i + e_j \}, \qquad
1\leq i<j \leq r ,
\end{eqnarray}
while for $\mathfrak{g}\simeq so(2r) $ have the form:
\begin{eqnarray}\label{eq:d1-so}
\Delta ^+_0 &=& \{ e_i - e_j\}, \qquad
\Delta ^+_1 = \{  e_i + e_j \}, \qquad 1\leq i<j \leq r.
\end{eqnarray}
The root vectors in the typical representation are given by:
\begin{equation}\label{eq:E_a}
E_{e_i-e_j} = E_{ij} -(-1)^{i+j} E_{\bar{j}\bar{i}}, \qquad
E_{e_i+e_j} = E_{i\bar{j}} -\epsilon _0(-1)^{i+j} E_{j\bar{i}},
\end{equation}
where $1\leq i \leq j \leq r $ and $\epsilon _0=\pm 1 $ as defined above.
Note that for $sp(2r) $ $\epsilon _0=-1 $ and eq. (\ref{eq:E_a}) provides
also the expressions for $E_{2e_j} $ by putting $i=j $; for $so(2r) $
$\epsilon _0=1 $ and putting $i=j $ in eq. (\ref{eq:E_a}) gives vanishing
result which is compatible with the fact that $2e_j $ are not  roots of
$so(2r) $.

The continuous spectrum of $L $ coinsides with the spectra of $L_\pm $.
There are no apriory restrictions on the locations of the discrete
eigenvalues of $L $ (\ref{eq:4.1}).

The construction of the fundamental analytic solution for the problem
(\ref{eq:4.1}) is rather tedious. Of all configurations for
the set of eigenvalues $\{ m_1,\dots, m_s\} $ we consider only the
generic one:
\begin{eqnarray}\label{eq:5b}
&& m_1>m_2 >\cdots > m_s>0.
\end{eqnarray}
Cases when subsets of $\{m_k\} $ are equal can be considered analogously.

Let us outline the construction of the FAS. The first peculiarity
is related to the fact that the spectrum multiplicity may vary with
$\lambda $, see the left panel of fig. ~\ref{fig:1}. This reflects
on the definition of the Jost solutions. Fot the case when all eigenvalues
of $Q_\pm $ arereal, i.e. of the type c) above we have:
\begin{eqnarray}\label{eq:9}
&& \psi (x,\lambda )
\mathop{\longrightarrow }\limits_{x\to \infty }
\psi _0(\lambda )e^{-iJ(\lambda )x} P(\lambda ), \qquad
\phi  (x,\lambda ) \mathop{\longrightarrow }\limits_{x\to -\infty }
\phi _0(\lambda )e^{-iJ(\lambda )x} P(\lambda ), \\
&& J(\lambda ) = \sum_{k=1}^{r} j_k(\lambda )H_k, \qquad
P(\lambda ) = \exp \left(2\pi i\sum_{k=1}^{r} P_k(\lambda )H_k \right), \\
&& j_k(\lambda )= \sqrt{\lambda ^2 -m_k^2},
\qquad  P_k(\lambda )= \theta (|\re \lambda |-m_k).\nonumber
\end{eqnarray}
The $x $-independent matrices $\psi _0(\lambda ) $ and $\phi _0(\lambda )
$ take values in the corresponding group $\mathfrak{G} $ and satisfy
\begin{equation}\label{eq:10}
U_+(\lambda ) \psi _0(\lambda )  = -\psi _0(\lambda ) J(\lambda ), \qquad
U_-(\lambda )\phi _0(\lambda )= -\phi _0(\lambda ) J(\lambda ),
\end{equation}
and are of the form:
\begin{eqnarray}\label{eq:11}
&&\psi _0(\lambda ) = \varphi _0^+ U_0(\lambda ) , \qquad
\phi _0(\lambda ) = \varphi _0^- U_0(\lambda ) , \nonumber\\
&& \varphi _0^\pm = \left( \begin{array}{cc} \underline{\varphi}_1^\pm &
0 \\ 0 & \underline{\varphi }_2^\pm \end{array}\right), \qquad
U_0(\lambda ) =  \left(\begin{array}{cc} \hat{S}_1 \underline{A} S_1 &
S_1 \underline{B} \\ \underline{B} S_1 & \underline{A} \end{array}\right),
\end{eqnarray}
Here the $r\times r $ matrices $\underline{\phi }_i^\pm $,
$\underline{A}$, $\underline{B} $, $S_1 $ are given by:

\begin{eqnarray}
&&\underline{A}_{kl} = \sqrt{{\lambda +j_k\over 2j_k}}\delta _{kl}, \qquad
\underline{B}_{kl} = \sqrt{{\lambda -j_k  \over 2j_k  }}\delta _{kl},
\qquad S_1 = \sum_{s=1}^{r}(-1)^{s+1} E_{s,r+1-s}, \nonumber\\
&& q_\pm q_\pm^\dag \underline{\varphi }_1^\pm =
-\underline{\varphi }_1^\pm \underline{m}^2, \qquad
q_\pm^\dag q_\pm \underline{\varphi }_2^\pm =
-\underline{\varphi }_2^\pm \underline{m}^2, \qquad
\underline{m}_{kl}= m_k\delta _{kl}.
\end{eqnarray}

Usually only the first and last columns of the Jost solutions
$\psi(x,\lambda ) $ and $\phi(x,\lambda ) $ are analytic in $\lambda $.
Nevertheless, applying the method in \cite{LOMI131} we can construct
fundamental solutions on each of the sheets of the $2^s $-sheeted
Riemannian surface, related to the set of roots $j_k(\lambda)$; each sheet
of this surface is determined by the set of signs of $\{\epsilon
_k(\lambda ):  \epsilon _k = \mbox{sign}\, \im j_k(\lambda )\} $. Let us
outline the construction of $\chi ^+(x,\lambda ) $ analytic in the region
of the sheet $\mathcal{S}_1 $ where:

\begin{equation}\label{eq:S1}
\mathcal{S}_1 \quad \colon \quad \im j_1(\lambda ) > \im j_2(\lambda ) >
\dots > \im j_r(\lambda ) > 0.
\end{equation}
First we construct the Gauss decomposition of the scattering matrix:
\begin{equation}\label{eq:T-gaus}
T(\lambda ) = T^-(\lambda ) D^+(\lambda ) \hat{S}^+(\lambda ),
\end{equation}
where
\begin{eqnarray}\label{eq:TDS}
T^-(\lambda ) = \exp \left( \sum_{\alpha >0}^{} \tau^-_\alpha (\lambda )
E_{-\alpha }\right), \qquad
S^+(\lambda ) = \exp \left( \sum_{\alpha >0}^{} \sigma ^+_\alpha (\lambda
) E_{\alpha }\right), \nonumber\\
D^+(\lambda ) = \exp \left( \sum_{k=1}^{r}
{2 \delta ^+_k(\lambda) \over (\alpha _k,\alpha _k)}H_{\alpha_k}\right),
\end{eqnarray}
where $H_{\alpha _k} $ are Cartan elements dual to the simple roots
$\alpha _k $  of $\mathfrak{g} $. Then the solution $\chi^+ (x,\lambda)$:
\begin{equation}\label{eq:13}
\chi  ^+ (x,\lambda ) = \psi (x,\lambda ) T^-D^+(\lambda )  = \phi
(x,\lambda )  S^+(\lambda )
\end{equation}
is analytic\footnote{More precisely it is not the functions
$\chi  ^+(x,\lambda ) $ that is analytic in $\lambda  $
but $\chi  ^+ (x,\lambda )e^{iJ(\lambda )x} $.} in $\lambda  $ on the
sheet ${\mathcal{S}}_1 $. Skipping the details of the proof which will be
published elsewhere we only give some additional usefull facts about $\chi
^+(x,\lambda ) $.

First, the function $D^+(\lambda ) $ is also analytic function of $\lambda
$ in $\mathcal{S}_1 $ which generates the integrals of motion for the
MNLS. Using the properties of the fundamental representations of the $C_r
$ and $D_r $ series we have:

\begin{equation}\label{eq:T-D-}
\langle \omega_j |T(\lambda )| \omega _j\rangle =
\langle \omega_j |D^+(\lambda )| \omega _j\rangle = \exp ((\omega _j,
\vec{\delta }^+(\lambda ))),
\end{equation}
where $\omega _j $ is the $j $-th fundamental weight of $\mathfrak{g} $
and $\vec{\delta }^+(\lambda ) = \sum_{k=1}^{r}\delta_k^+(\lambda )e_k $.
Note that the simple roots $\alpha _k $ and the fundamental weights
$\omega _j $ satisfy the relation $2(\omega _j,\alpha _k)/(\alpha
_k,\alpha _k)=\delta _{jk} $ More specifically for our examples we have:
\begin{eqnarray}\label{eq:dp-sp}
\delta _1^+(\lambda ) = \ln T_{11}(\lambda ), \qquad
\delta _2^+(\lambda ) = \ln \left\{ \begin{array}{cc} 1 & 2 \\ 1 & 2
\end{array}\right\}_ {T(\lambda )},
\end{eqnarray}
for $sp(4) $ and
\begin{eqnarray}\label{eq:dp-so}
&& \delta _1^+(\lambda ) = \ln T_{11}(\lambda ), \qquad
\delta _2^+(\lambda ) = \ln \left\{ \begin{array}{cc} 1 & 2 \\ 1 & 2
\end{array}\right\}_ {T(\lambda )}, \\
&&\delta _3^+(\lambda ) = \ln \left\{ \begin{array}{ccc} 1 & 2 &3 \\ 1 & 2
& 3 \end{array}\right\}_ {T(\lambda )} - \delta _4^+(\lambda ),
\qquad  \delta _4^+(\lambda ) = {1  \over 2 } \ln \left\{
\begin{array}{cccc} 1 & 2 & 3 & 4\\ 1 & 2 & 3 & 4\end{array}\right\}_
{T(\lambda )} ,\nonumber
\end{eqnarray}
for $so(8) $.
Here by $\left\{ \begin{array}{ccc} 1 & \dots & k \\ 1 & \dots & k
\end{array}\right\}_ {T(\lambda )} $ we denote the upper principal minor
of order $k $ of the scattering matrix $T(\lambda ) $.

Note that due to the orthogonal symmetry inherent in $D_4 $ all functions
$\exp (\delta _k^+(\lambda )) $ are polynomial expressions in terms of the
matrix elements of $T(\lambda ) $.

Fundamental analytic solutions can be constructed in analogous way on each
of the sheets of the Riemannian surface.

\section{Examples of MNLS with CBC on  symmetric spaces}\label{ssec:5.1}


The requirement that $U(x,t,\lambda ) $ and $V(x,t,\lambda ) $ belong to $
\mathfrak{g} $ (see eq. (\ref{eq:Lie})) can be formulated as the reduction
condition:
\begin{eqnarray}\label{eq:g}
S_0^{-1}U^T(x,t,\lambda )S_0 = -U(x,t,\lambda ),&  \qquad S_0^{-1}J S_0
=-J,\nonumber\\
S_0^{-1} V^T(x,t,\lambda )S_0 = -V(x,t,\lambda ),
\end{eqnarray}
which has trivial action on $\lambda  $. Such reduction imposes
restrictions only on the coefficients of $Q(x,t) $ so that for $C_r $ we
can put:
\begin{equation}\label{eq:Cr}
Q(x,t) = \sum_{i<j}^{}  (q_{ij} E_{e_i+e_j} + r_{ij}E_{-e_i-e_j}) +
{1  \over 2 }\sum_{j=1}^{r} (q_{i} E_{2e_j} + r_{i} E_{-2e_j} ),
\end{equation}
while in the $D_r $-case we have;
\begin{equation}\label{eq:Dr}
Q(x,t) = \sum_{i<j}^{}  (q_{ij} E_{e_i+e_j} + r_{ij} E_{-e_i-e_j}) ,
\end{equation}
In the typical representations of $C_r $ and $D_r $ these choices for
$Q(x,t) $ have always the block structure shown in (\ref{eq:Q}). In the
case of $\mathfrak{g}\simeq sp(4) $ the blocks $\q $ and $\br $ are
parametrized by three functions each:
\begin{equation}\label{eq:sp4}
\q(x,t) = \left( \begin{array}{cc} q_{12} & q_1 \\ q_2& q_{12}
\end{array}\right), \qquad \br(x,t) = \left( \begin{array}{cc} r_{12} &
r_2 \\ r_1 & r_{12} \end{array}\right),
\end{equation}
while for $\mathfrak{g}\simeq so(8) $ they contain six independent
functions each:
\begin{equation}\label{eq:so8}
\q(x,t) = \left( \begin{array}{cccc} q_{14} & q_{13}& q_{12} & 0 \\
q_{24} & q_{23}& 0 & q_{12}  \\ q_{34} &0 & q_{23}& -q_{13}  \\
0 &q_{34} & -q_{24} & q_{14}  \end{array}\right), \qquad
\br(x,t) = \left( \begin{array}{cccc} r_{14} & r_{24}& r_{34} & 0 \\
r_{13} & r_{23}& 0 & r_{34}  \\ r_{12} &0 & r_{23} & -r_{24}  \\
0 &r_{12} & -r_{13} & r_{14}  \end{array}\right),
\end{equation}
The corresponding sets of MNLS eqs. (\ref{eq:3.1}) for these two choices
of $Q(x,t) $ and for VBC were first derived in \cite{ForKu*83}. For CBC
with $\br=\q^\dag $ they take the form (\ref{eq:1}) with the additional
linear in $\q $ terms ensuring regular behavior for $t\to\pm\infty  $.

Note that reducing $Q(x,t) $ to take values in $\mathfrak{g} $ then we
naturally have that $\psi _0(\lambda ) $ and $\phi _0(\lambda ) $ take
values in the corresponding group $\mathfrak{G} $.

\subsection{Spectral properties of $sp(4) $--MNLS with CBC}\label{sec:C2}

As mentioned in Section~3, the continuous spectrum of the GZS system
(\ref{eq:4.1}) is determined by the set of eigenvalues $\{ \nu _j,
j=1,2\} $ of the matrices $q_+r_+=q_-r_- $. These eigenvalues for $Q(x,t)
$ given by eqs. (\ref{eq:Cr}), (\ref{eq:sp4}) with $r=2 $ satisfy the
characteristic equation:
\begin{equation}\label{eq:chpo}
\nu ^2 -K_0 \nu +K_1=0, \qquad K_0={1 \over 2}\tr Q_\pm^2, \qquad K_1=\det
Q_\pm.
\end{equation}
and determine the end points of the spectrum.
If we impose on $Q(x,t) $, and consequently on $Q_\pm $
the involution ($\bbbz_2 $-reduction):
\begin{equation}\label{eq:Z2-r}
B_1^{-1} Q^\dag B_1 = Q, \qquad B_1 =\diag(1,\epsilon,\epsilon ,1),
\qquad \epsilon =\pm 1.
\end{equation}
which in components takes the form:
\begin{equation}\label{eq:Z2-r2}
r_1=\epsilon q_1^*, \qquad r_2=q_2^*, \qquad r_3=q_3^*.
\end{equation}
Then the coefficients $K_0 $ and $K_1 $ equal:
\begin{equation}\label{eq:K-i}
K_0 = 2\epsilon |q_{1}^\pm|^2 + |q_{2}^\pm|^2 + |q_{3}^\pm|^2, \qquad
K_1 = |(q_{1}^\pm)^2 + q_{2}^\pm q_{3}^\pm  |^2
\end{equation}
We have three possibilities for the roots $\nu _1,\nu _2 $  of eq.
(\ref{eq:chpo}) depending on the sign of the discriminant:
\begin{equation}\label{eq:discr}
D={1  \over 4 } K_0^2 - 4K_1.
\end{equation}

\begin{description}

\item[a) $D>0 $,] i.e. the roots $\nu _{1}>\nu _2 $ are different and
real. The continuous spectrum of $L $ fills up two pairs of rays on the
real axis $|\re \lambda| >\nu _1$ and $ |\re \lambda| > \re \nu _2$;

\item[b) $D=0 $, ] i.e. the roots $\nu _1=\nu _2 $; the two pairs of rays
in a) now coincide; the total multiplicity of the spectrum is $4 $;

\item[c) $D<0 $,] i.e. the roots $\nu _j $ are complex-valued and $\nu
_1=\nu _2^* $; The continuous spectrum of $L $ fils up two branches of
two-fold spectrum along the hyperbola's arcs $\re \lambda\, \im
\lambda = \re \nu _k \im \nu _k$, see the right panel of fig.~\ref{fig:1};

\end{description}

In the generic case there are no apriory limitations as to the positions
of the discrete eigenvalues. Such may come up if we consider potentials
$Q=-Q^\dag $; then the GZS system become equivalent to a formally
self-adjoint linear problem whose spectrum should be confined to the real
$\lambda  $-axis only.  The formal self-adjointness takes place for
$\epsilon =1$.

\subsection{Spectral properties of $so(8) $-MNLS with CBC}\label{sec:D4}

The characteristic equation for $q_\pm r_\pm $ takes more simple form:
\begin{equation}\label{eq:chp}
\det (q_\pm r_\pm -\nu )= (\nu ^2 -K_0\nu +K_1)^2,
\end{equation}
where the coefficients $K_j $ now are given by:
\begin{eqnarray}\label{eq:K-j}
K_0&=& {1 \over 2}\tr (q_\pm r_\pm) = \sum_{1\leq i<j\leq 4}
q_{ij}^{\pm} r_{ij}^{\pm}, \\
K_1&=& (\det (q_\pm r_\pm))^{1/2}= (
q_{13}^{\pm} q_{24}^{\pm} - q_{34}^{\pm} q_{12}^{\pm}-q_{23}^{\pm}
q_{14}^{\pm} ) (r_{13}^{\pm} r_{24}^{\pm} - r_{34}^{\pm}
r_{12}^{\pm}-r_{23}^{\pm} r_{14}^{\pm} ).\nonumber
\end{eqnarray}
An involution of the type (\ref{eq:Z2-r}) gives $r_{ij}=\epsilon
_i\epsilon _jq_{ij}^* $ with $\epsilon _j=\pm 1 $ and makes the
coefficients $K_0 $, $K_1 $ real. Besides now each of the eigenvalues
$\nu _j $, $j=1,2 $ is two-fold.  Again we have the three possibilities
depending on the value of $D $; the only difference is that the
multiplicity of each of the branches is $4 $.
This imposes certain symmetry on the locations of the eigenvalues of $\nu
_j $ which in fact determine the end-points of the continuous spectra of $
L $.


\section{Hamiltonian properties}\label{sec:}

The invariants of the transfer matrix $T(\lambda ) $ such as, e.g. its
principal minors generate integrals of motion, i.e. if all $\nu _j $ are
different we have only $r $ independent series of conserved quantities.
Let us briefly outline the methods of deriving of these integrals as
functionals of the potential $Q $. As starting relation here we consider
the Wronskian relation, generalizing the one derived in \cite{12} for the
scalar case $s=1 $:
\begin{eqnarray}\label{eq:33}
&& \left.\tr \left[ i(\chi ^+)^{-1} {d\chi ^+  \over d\lambda  }
C - {dJ(\lambda )  \over d\lambda  } x C\right] \right|_{x=-\infty
}^{\infty } \nonumber\\
&& = i \sum_{k=1}^{r} {d \delta _k^+\over d\lambda  } \tr (H_kC) +
i \tr \left( \hat{\psi }_0(\lambda ) {d\psi _0  \over d\lambda  } C
- \hat{\phi }_0(\lambda ) {d\phi _0  \over d\lambda  } C  \right) \\ && =
\int_{-\infty }^{\infty } dx\, \tr \left[ \sigma _3 R_C(x,t,\lambda ) -
\lambda J^{-1}(\lambda ) C \right], \nonumber
\end{eqnarray}
where $C $ is a constant element of $\mathfrak{h} $ and  $R_C(x,t,\lambda )
= \chi ^+C(\chi ^+)^{-1}(x,t,\lambda ) $ is a natural generalization of the
diagonal of the resolvent of the system (\ref{eq:4.1}). It satisfies the
equation:
\begin{equation}\label{eq:34}
i {dR_C  \over  dx} + \left[ Q(x,t) - \lambda J, R_C \right]
=0, \qquad \lim_{x\to\infty } R_C(x,t,\lambda ) = \psi _0(\lambda ) C\psi
_0^{-1}(\lambda ).
\end{equation}
Eq. (\ref{eq:34}) allows one to derive the recurrent relations for
evaluating the expansion coefficients
\begin{equation}\label{eq:R-C}
R_C(x,t,\lambda ) = C_0 + \sum_{k=1}^{\infty } R_k\lambda ^{-k}(x,t) ,
\qquad \psi_0 C\psi_0^{-1}(\lambda ) = C_0 + \sum_{k=1}^{\infty }
C_k\lambda^{-k}.
\end{equation}

The trace identities for the MNLS type equations with CBC can be derived
by inserting the asymptotic expansions of $R_C(x,\lambda ) $ and $\delta
_k^+(\lambda )$:
\begin{equation}\label{eq:as-ex}
\delta _k^+(\lambda ) = \sum_{p=1}^{\infty } I_{p}^{(k)}\lambda ^{-p}
\end{equation}
in both sides of eq. (\ref{eq:33}) and equating the corresponding
coefficients of $\lambda ^{-p} $.

Here we write down the first three of the local integrals of motion coming
from the principal series with $C=J $:
\begin{eqnarray}\label{eq:35}
&& \H_k= \int_{-\infty }^{\infty } dx\, \tr \left(\sigma _3 R_{k+1}(x,t) -
C_{k+1}\right), \qquad
\H_1 = {1  \over 2 }\int_{-\infty }^{\infty } dx\, \tr
(q q^\dag(x,t) - \bar{\mu}  ), \nonumber\\
&& \H_2 = {i  \over 4 }  \int_{-\infty }^{\infty } dx\, \tr
(q_x q^\dag - qq_x^\dag ), \qquad \bar{\mu } = q_+ q_+^\dag,\\
&& \H_3 = {3  \over 8 }  \int_{-\infty }^{\infty } dx\, \tr \left[
q_x q_x^\dag + (qq^\dag (x,t))^2 - \bar{\mu }^2 )\right].\nonumber
\end{eqnarray}
The correct use of the Wronskian relation (\ref{eq:33}) allowed us to
derive renormalized integrals of motion, i.e. ones that converge for
$Q(x,t)\in \mathcal{M}$.

However among the integrals in this series one can not find the
Hamiltonian of the MNLS (\ref{eq:1}). In order to obtain the Hamiltonian
we need to regularize these integrals. By regularized integral we mean one
whose gradient $\delta \H_k/\delta Q^T(x,t) $ vanishes for both
$x\to\infty $ and $x\to -\infty  $. This can be done by considering
additional series  of integrals, which generically have non-local
densities. Fortunately among the simplests of them one may find local
ones. For example, the first integral from the series with $C$ choosen to
be  $C^{(l)} = \sum_{k=1}^{r} m_{k}^{2l} H_k$ with $1\leq l\leq r $, is
local and has the form:
\begin{equation}\label{eq:36}
\tilde{\H}_1^{(l)} = {1  \over 4 } \int_{-\infty }^{\infty } dx\, \tr
\left[ qq^\dag(x,t) \bar{\mu }^l + q^\dag q(x,t) \mu ^l - 2 \mu ^{l+1}
\right], \qquad \mu =q_+^\dag q_+.
\end{equation}
Note, that $\tilde{\H}_1^{(l)} $ is nontrivial, i.e. does not reduce to
$\H_1 $ only if $s\geq 2 $, $\nu_1\neq \nu_2 \neq \dots $.  Using it we
can check the validity of
\begin{equation}\label{eq:37}
\H_{\mbox{\scriptsize MNLS}}= {8  \over 3 } \H_3 - 4 \tilde{\H}_1^{(1)} =
\int_{-\infty }^{\infty } dx\, \tr \left[ q_x q_x^\dag + (qq^\dag (x,t) -
\bar{\mu })^2 \right].
\end{equation}

In analyzing the Hamiltonian properties of the MNLS with CBC we will make
use also of the classical $r $-matrix approach, see \cite{4,FaTa}. It
allows one to write down in compact form the Poisson brackets of the
transfer (monodromy) matrix. Since our problem is ultra-local in
the terminology of \cite{4,FaTa} then the definition of $r $ is
independent on the boundary conditions. Taking into account the results of
\cite{ForKu*83} we find
\begin{equation}\label{eq:r-mat}
r(\lambda ,\mu ) = {1  \over \lambda -\mu  } \left( \sum_{\alpha \in
\Delta _0^+ \cup \Delta _1^+} (E_{\alpha } \otimes E_{-\alpha } +
E_{-\alpha }\otimes E_{\alpha }) + \sum_{j=1}^{r} H_j\otimes H_j \right),
\end{equation}
where $E_{\alpha } $ and $H_j $ are the Cartan-Weyl generators of
the corresponding simple Lie algebra $\mathcal{g} $, see \cite{Helg}.  In
order to derive the Poisson brackets for the MNLS on the whole axis with
CBC we need to take into account the corresponding oscillations coming
from the Jost solutions.

Skipping the details we just write down the  expressions for the Poisson
brackets between the matrix elements of $T(\lambda ) $:
\begin{eqnarray}\label{eq:38}
&& \left\{ T(\lambda ) \otimescomma T(\mu ) \right\} = r_+(\lambda ,\mu )
 T(\lambda ) \otimes T(\mu ) - T(\lambda )\otimes T(\mu ) r_-(\lambda
,\mu ),\\
&& r_\pm(\lambda ,\mu ) = \lim_{x\to\pm\infty } \tau_\pm^{-1}(x,\lambda
,\mu ) r(\lambda ,\mu ) \tau_\pm(x,\lambda ,\mu ),\nonumber\\
&& \tau_+(x,\lambda ,\mu ) = \psi _0(\lambda )e^{-iJ(\lambda )x} \otimes
\psi _0(\mu ) e^{-iJ(\mu )x}, \nonumber\\
&& \tau_-(x,\lambda ,\mu ) = \phi _0(\lambda )e^{-iJ(\lambda )x} \otimes
\phi _0(\mu ) e^{-iJ(\mu )x}, \nonumber
\end{eqnarray}
where $\{T(\lambda )\otimescomma T(\mu )\}_{ij,kl} \equiv \{T_{ij}(\lambda
),T_{kl}(\mu )\} $.

Similar problem for the scalar case was analyzed in \cite{12,FaTa}.  For
$\mathfrak{g}\simeq C_r $, or $D_r $ the explicit form of $r_\pm(\lambda
,\mu ) $ is rather complicated and will be published elsewhere. An
important and difficult problem here is to take correctly into account the
the threshold singularities of $T_{kl}(\lambda ) $ of the form
$j_k^{-1}(\lambda ) $ at the end points of the continuous spectrum.

An important consequence of (\ref{eq:38}) are the involution properties of
$\delta _k^+(\lambda ) $:
\begin{equation}\label{eq:Invol}
\{ \delta _k^+(\lambda ), \delta _j^+(\mu )\} =0,
\end{equation}
for all values of $1\leq i,j\leq r $ and $\lambda  $ and $\mu  $ taking
values on the continuos spectrum of $L $. From (\ref{eq:Invol}) there
follows that $\{I_{p}^{(k)}, I_{s}^{(l)}\}=0 $ for all positive values of
$p $ and $s $, and for all $1\leq k, l\leq r $. A consequence of eq.
(\ref{eq:Invol}) is the  involutivity of the integrals of the principal
series $\{ \H_k , \H_p \}=0$. This is a necessary condition in proving the
complete integrability of the MNLS equations with CBC; other difficulties
in proving it are outlined in \cite{11}.

Of course the rigorous proof of the complete integrability and the
derivation of the basic properties of the MNLS equations must be
based on the completeness relation of the relevant `squared solutions' of
$L $.  For the single component NLS such relation has been proposed in
\cite{Konotop}; for the multicomponent systems this is still open
question.

\section{Discussion}\label{sec:4}

The $sp(4) $ MNLS can be viewed as a special reduction from the $su(4)$
one. Recently it was discovered that the $sp(4) $ MNLS with vanishing BC
has important applications to BEC \cite{Wada}. This enhances the interest
to the MNLS type models. In particular it will be important  to work out
the dressing Zakharov-Shabat procedure not only for MNLS with vanishing
boundary conditions, but also for non-vanishing BC. Some steps in this
directions have been reported in \cite{LOMI131} for the
$su(n+m)/s(u(n)\otimes su(m)) $ case. Deriving the dressing factors for
the MNLS for the  symmetric spaces of types $C.II $- and $D.III $- requires
substantial changes even for vanishing BC; doing the same for CBC is still
a bigger chalenge.

Similar methods can be applied to the analysis of the $N $-wave type
equations with CBC, see \cite{Dokt}.

\section*{ Acknowledgements} This work has been  supported by the National
Science Foundation of Bulgaria, contract No.  F-1410.

\end{document}